\documentclass{elsart}
\usepackage{amssymb,amsfonts,graphicx}
\usepackage{rotating}
\usepackage{subfigure}
\usepackage{dcolumn}
\usepackage{bm}
\usepackage{epsfig}

\begin{document}

\begin{frontmatter}

\title{Information Flow between Composite Stock Index and Individual Stocks}

\author[korea1,korea2]{Okyu Kwon},
\author[korea1]{Jae-Suk Yang\corauthref{cor1}}
\ead{mathphy@korea.ac.kr}
\corauth[cor1]{Corresponding author.}
\address[korea1]{Department of Physics, Korea University, Seoul 131-701, Korea}
\address[korea2]{National Creative Research Initiative Center for
Neuro-dynamics, Korea University, Seoul 136-701, Korea}

\begin{abstract}

We investigate the strength and the direction of information
transfer in the U.S. stock market between the composite stock
price index of stock market and prices of individual stocks using
the transfer entropy. Through the directionality of the
information transfer, we find that individual stocks are
influenced by the index of the market.

\end{abstract}


\end{frontmatter}


\section{Introduction}

Recently, economy has become an active research area for
physicists. Physicists have attempted to apply the concepts and
methods of statistical physics, such as the correlation function,
multifractal, spin models, complex networks, and information
theory to study economic problems
\cite{arthur,mategna,bouchaud,mandel,kullmann,giada,eguiluz,krawiecki,takaishi,kaizoji02,jbpark,jwlee,wsjung1,wsjung2,matal,yang,schae,jsyang2}.

From the economic system, many empirical data reflecting the
economic conditions can be obtained. Among them the time series of
composite stock price index is one of the best data reflecting
economics conditions well. The index data is used to analyze and
predict the perspective of markets. The scientific interest in
studying financial markets stems from the fact that there is a
large amount of reasonably well defined data.

Information is an important keyword in analyzing the market or in
estimating the stock price of a given company. It is quantified in
rigorous mathematical terms \cite{shannon}, and the mutual
information, for example, appears as meaningful choice replacing a
simple linear correlation even though it still does not specify
the direction. The directionality, however, is required to
discriminate the more influential one between correlated
participants, and can be detected by the transfer entropy
\cite{schreiber}.

In many case, traders in the stock market refer to the index to
invest in stocks. Therefore, we can guess that prices of stocks is
affected by the composite stock index of the market. However, No
attempt to measure the influence of index quantitatively has been
accomplished, while it is found evident that the interaction
therein is highly nonlinear, unstable, and long-ranged from many
previous research on econophysics using financial time series.
Schreiber \cite{schreiber} introduced the transfer entropy which
measures dependency in time between two variables. We focus
quantitatively on the direction of information flow between the
index data and the price of individual companies using the method
of the transfer entropy. This concept of the transfer entropy has
been already applied to the analysis of financial time series by
Marschinski and Kantz \cite{marschinski}. They calculated the
information flow between the Dow Jones and DAX stock indexes and
obtained conclusions consistent with empirical observations. While
they examined interactions between two huge markets, we construct
its internal structure between stock index and individual stocks.

\section{Transfer entropy}

The transfer entropy which measures directionality of variable
with respect to time has been recently introduced by Schreiber
\cite{schreiber} based on the probability density function (PDF).
Let us consider two discrete and stationary process, $I$ and $J$.
The transfer entropy relates $k$ previous samples of process $I$
and $l$ previous samples
 of process $J$ is defined as follows:
\begin{equation}
T_{J \rightarrow I} = \sum p(i_{t+1}, i_t^{(k)}, j_t^{(l)}) \log
\frac{p( i_{t+1} \mid i_t^{(k)}, j_t^{(l)})}{p( i_{t+1} \mid
i_t^{(k)})},
\end{equation}
where $i_t$ and $j_t$ represent the discrete states at time $t$ of
$I$ and $J$, respectively. $i_t^{(k)}$ and $j_t^{(l)}$ denotes $k$
and $l$ dimensional delay vectors of two time consequences $I$ and
$J$, respectively. The joint PDF $p(i_{t+1}, i_t^{(k)},
j_T^{(l)})$ is the probability that the combination of $i_{t+1}$,
$i_t^{(k)}$ and $j_t^{(l)}$ have particular values. The
conditional PDF $p(i_{t+1} \mid  i_t^{(k)}, j_t^{(l)})$ and $p(
i_{t+1} \mid i_t^{(k)})$ are the probability that $i_{t+1}$ has a
particular value when the value of previous samples $i_t^{(k)}$
and $j_t^{(l)}$ are known and $i_t^{(k)}$ are known, respectively.

The transfer entropy with index $J \rightarrow I$ measures how
much the dynamics of process $J$ influences the transition
probabilities of another process $I$. The reverse dependency is
calculated by exchanging $i$ and $j$ of the joint and conditional
PDFs. The transfer entropy is explicitly asymmetric under the
exchange of $i_t$ and $j_t$. It can thus give the information
about the direction of interaction between two time series.

The transfer entropy is quantified by information flow from $J$ to
$I$. The transfer entropy can be calculated by subtracting the
information obtained from the last observation of $I$ only from
the information about the latest observation $I$ obtained from the
last joint observation of $I$ and $J$. This is the main concept of
the transfer entropy. Therefore, the transfer entropy can be
rephrased as
\begin{equation}
T_{J \rightarrow I} = h_I(k) - h_{IJ}(k,l),\label{eq:TE}
\end{equation}
where
\begin{eqnarray}
h_I(k) & = & -\sum p(i_{t+1}, i_t^{(k)}) \log p( i_{t+1} \mid i_t^{(k)}) \\
h_{IJ}(k,l) & = & -\sum p(i_{t+1}, i_t^{(k)}, j_t^{(l)}) \log p(
i_{t+1} \mid i_t^{(k)}, j_t^{(l)}).
\end{eqnarray}

\section{Empirical data analysis}

We analyze daily records of the S\&P 500 index (GSPC), Dow Jones
index (DJI) and stock price of selected 125 individual companies.
The dataset consists of about 4,000 simultaneously recorded data
points during the period June 1, 1983 to May 31, 2007. We use
logarithmic price difference as follows:
\begin{equation}
x_n \equiv \ln(S_n) - \ln(S_{n-1}),
\end{equation}
where $S_n$ means index or stock price of {\it n}-th trading day.
The first step in analysis for the transfer entropy is to
discretize the time series by some coarse graining. Quite often,
statistical studies which use the entropy assume that the
variables of interest are discrete, or may be discretized in some
straightforward manner. We partitioned the real value $x_n$ into
discretized price change $A_n$. In the concrete, $A_n=0$ for $x_n
\leq -d/2 $ (decrease), $A_n=1$ for $-d/2 < x_n < d/2$
(intermediate), $A_n=2$ for $x_n \geq d/2$ (increase) are chosen.

\begin{figure}[tbph]
\begin{center}
\mbox{
         {\scalebox{0.30}
       {\hspace{0mm}\epsfig{file=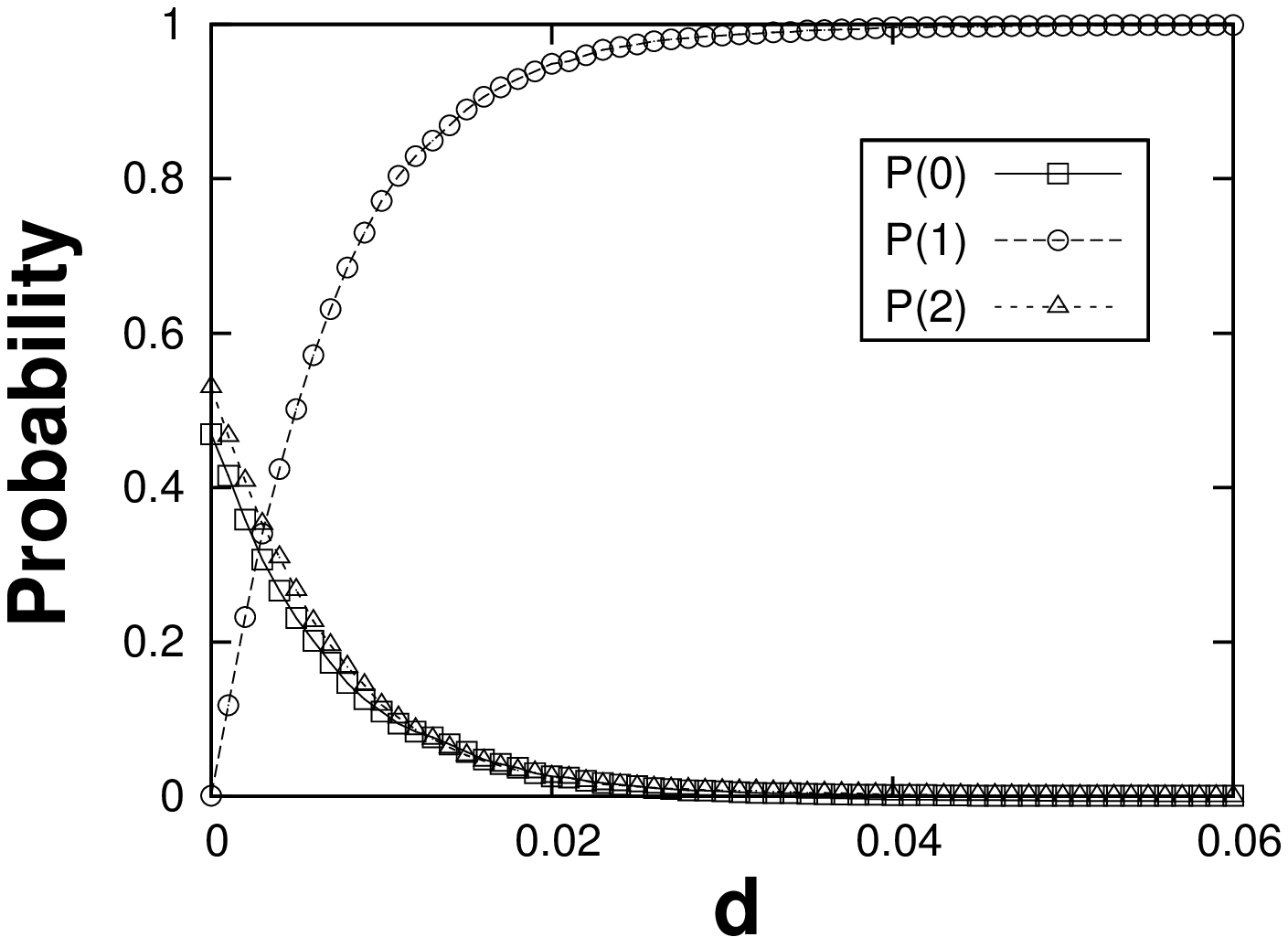, angle=0}} }
         {\scalebox{0.30}
       {\hspace{0mm}\epsfig{file=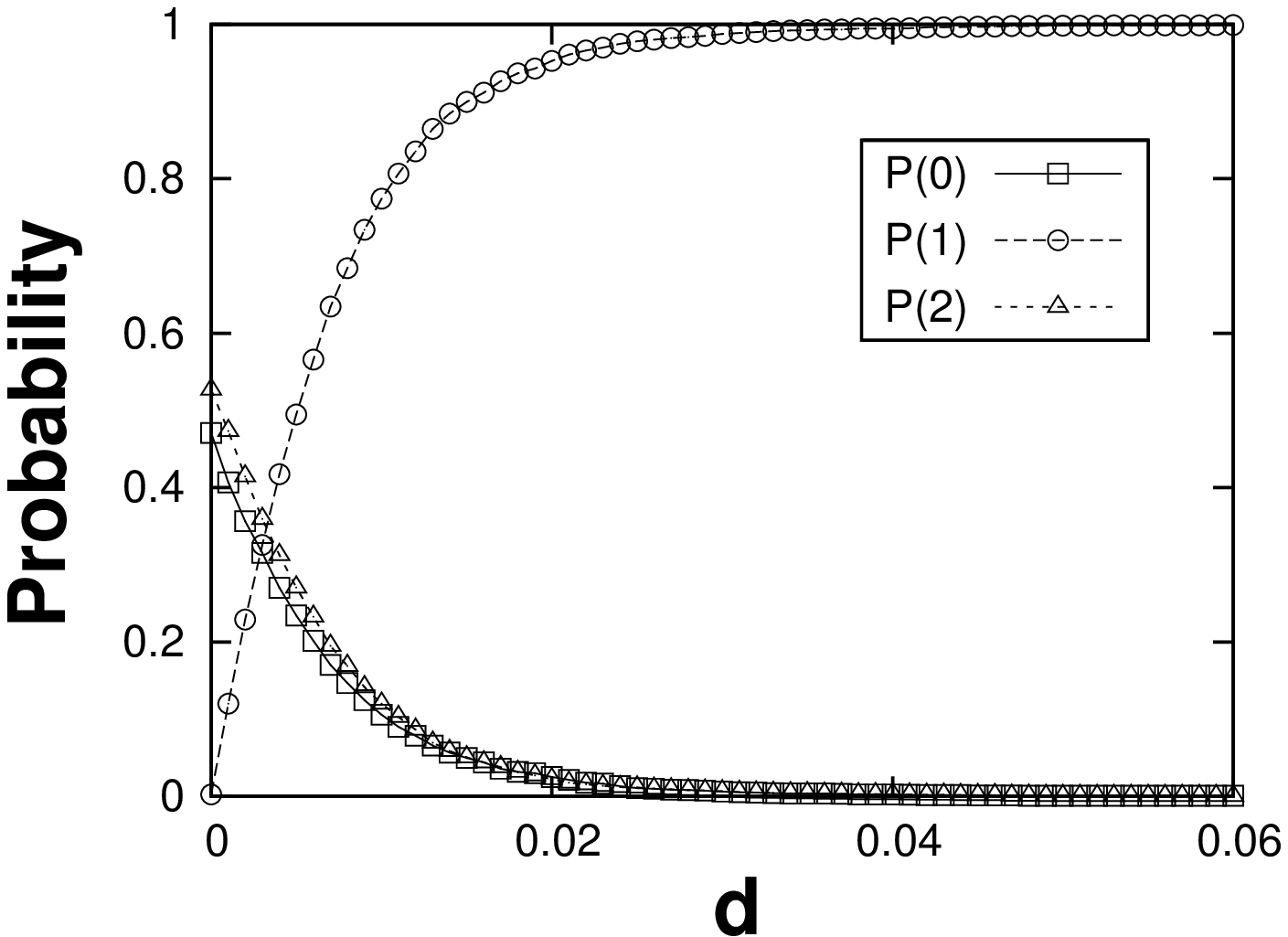, angle=0}} }
         {\scalebox{0.30}
       {\hspace{0mm}\epsfig{file=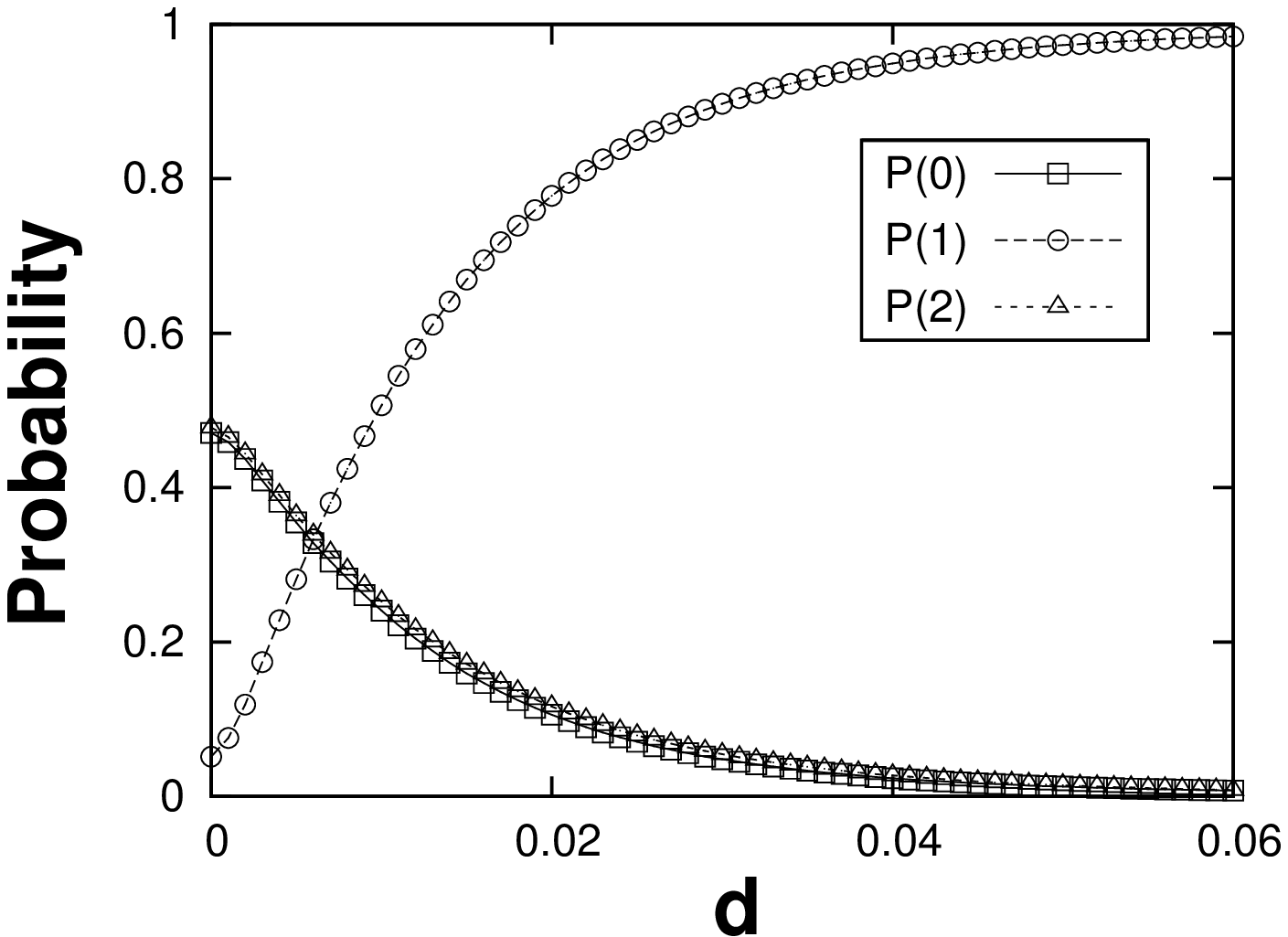,angle=0}} }
}
\end{center}
\vspace{0mm} \caption{Probability of states for (a) the GSPC, (b)
the DJI, and (c) individual stocks as a function of $d$, where (c)
is the average probability for all individual stocks. }
\label{fig:pstate}
\end{figure}

When data is discretized, it is important to determine the size of
$d$ because probability of each state is varied by $d$. In case of
very small $d$, most of return value is belonged not to the
intermediate state but to the increase or decrease states.
Therefore, the data can be regarded as two-states practically.
Also, when $d$ is very high, the greater part of return is fallen
under intermediate state. So data is able to be considered in
one-state system. As the value of $d$, the range of intermediate
state, is changed, the probability of each state is varied. Fig.
\ref{fig:pstate} represents the probability of each state. The
probabilities of increase and decrease states are almost same.
Therefore, the probability of intermediate state increases as $d$
is increasing, while those of increase and decrease are reduced.
Around $d=0.003$, the probabilities of three states are
approximately same for both of composite stock index. On the other
hand, individual stocks represent the same probability at
$d=0.006$. The reason, why the value of $d$ which makes the same
probability for composite stock index is not same to that for
individual stock prices, is that index usually does not change its
value abruptly in a day compare with individual stocks, because
composite stock index is average or weighted average of individual
stock prices.

\begin{figure}[tbph]
\begin{center}
\mbox{
                {\scalebox{0.35}
       {\hspace{0mm}\epsfig{file=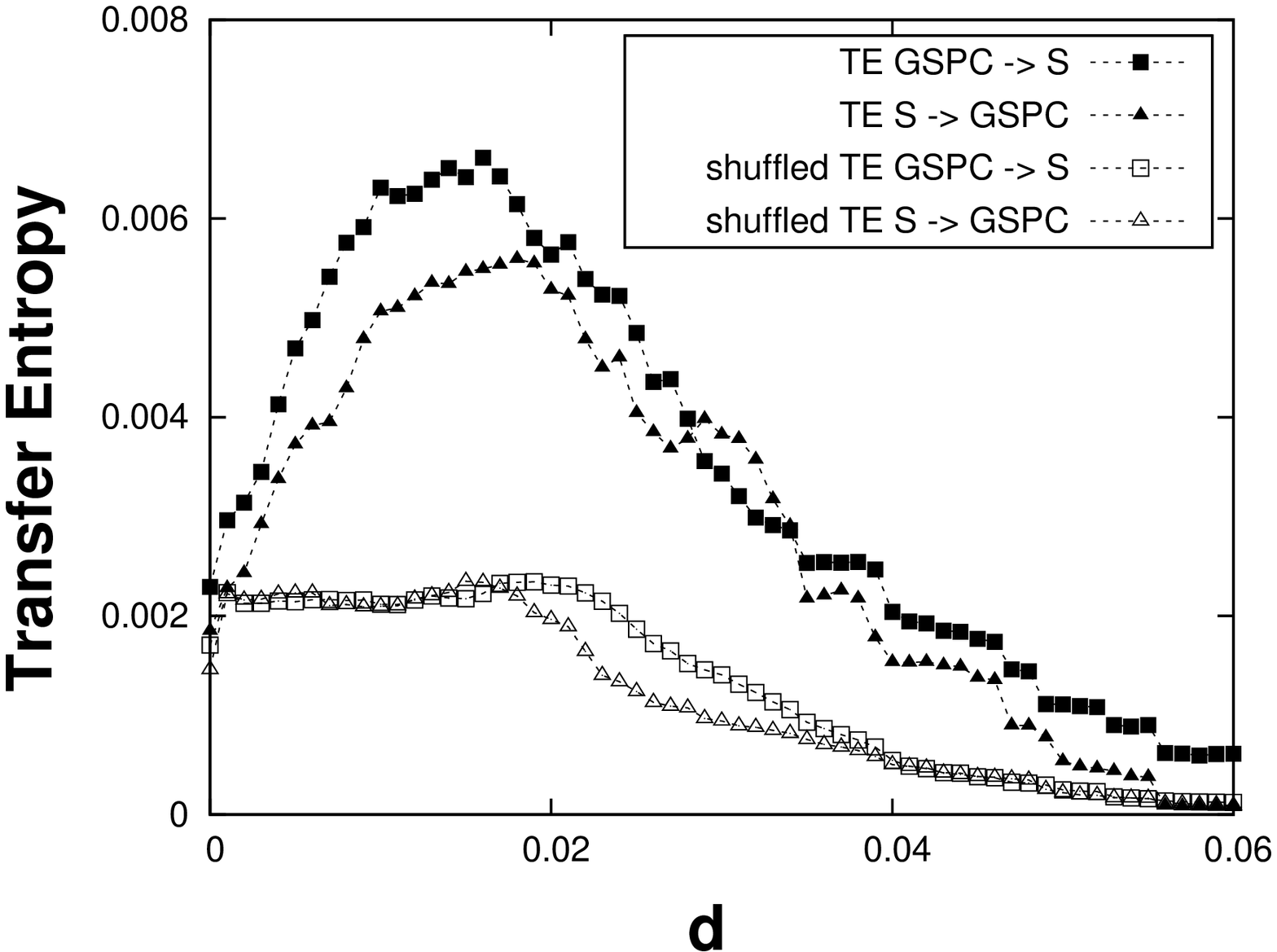, angle=0}} }
         {\scalebox{0.35}
       {\hspace{0mm}\epsfig{file=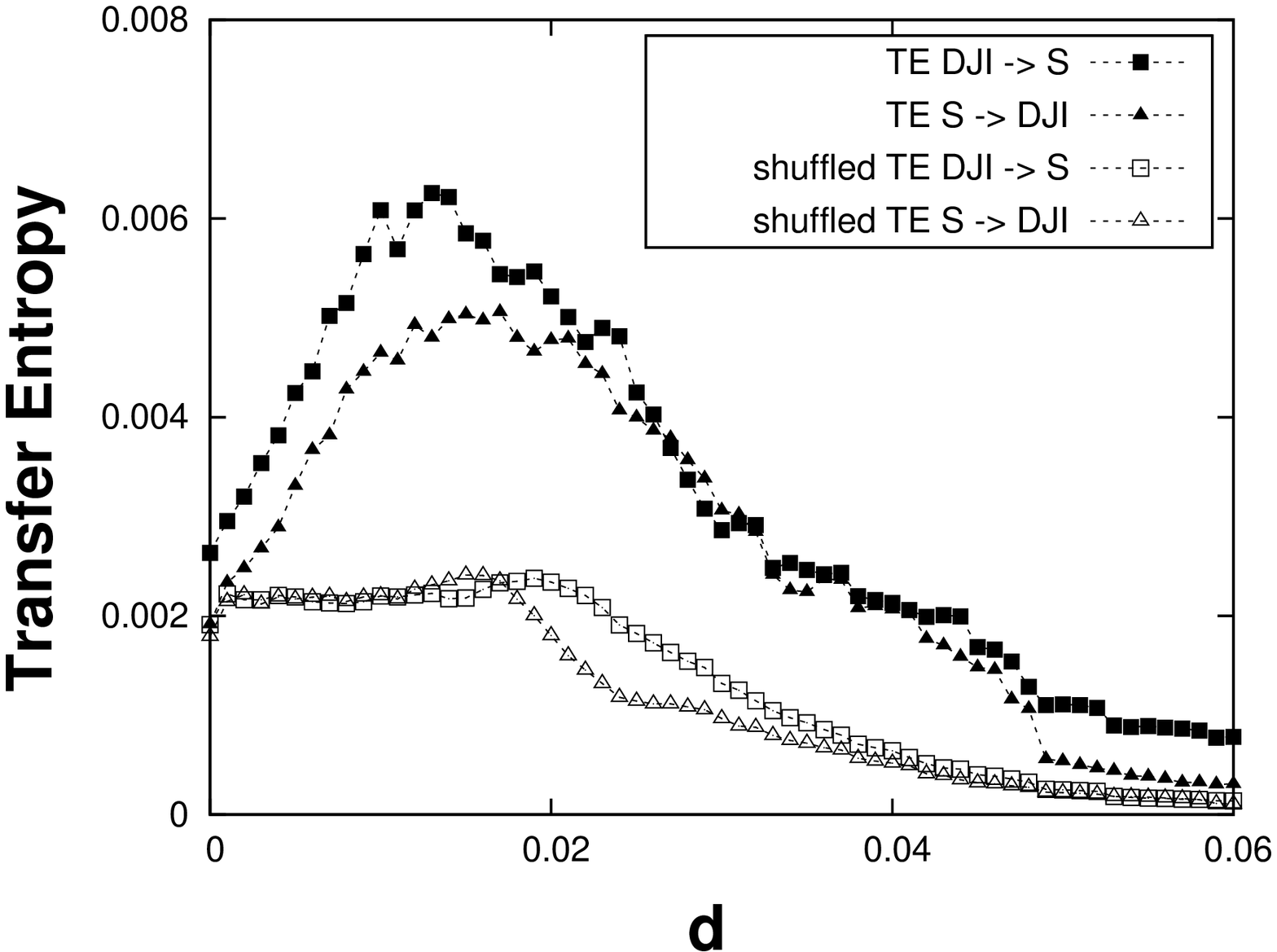, angle=0}} }
}
\end{center}
\vspace{0mm} \caption{Mean value of the transfer entropy for (a)
the GSPC and (b) the DJI as a function of $d$: $\blacksquare$ for
$T_{I \rightarrow S}$, $\blacktriangle$ for $T_{S \rightarrow I}$,
$\square$ for $T^{\rm shuffle}_{I \rightarrow S}$, and
$\vartriangle$ for $T^{\rm shuffle}_{S \rightarrow I}$.}
\label{fig:meanTE-d}
\end{figure}

Fig. \ref{fig:meanTE-d} shows the mean value of the transfer
entropy between composite stock index ($I$) and price of
individual stocks ($S$) for the GSPC and the DJI as a function of
$d$ with $k=1$ and $l=1$. The transfer entropy from the stock
index to the stock prices, $T_{I \rightarrow S}$, is almost higher
than that from the stock prices to the stock index, $T_{S
\rightarrow I}$. At $d=0$, discretized data is fallen into
two-states because the intermediate state is disappeared.
Therefore, it has smaller value of the transfer entropy compared
with that for three states. As $d$ is increasing, the number of
state turns to three, and the transfer entropy is maximized around
$d=0.015$.
Above $d$ which makes maximized the transfer entropy, the larger
$d$, the larger probability of the intermediate state. Moreover,
above about 0.02, $P(1)$ for the index goes close to 1. Therefore,
the transfer entropy is deceasing and finally goes to 0 because
all data is fallen into the intermediate state at very large $d$.

Open squares ($\square$) and triangles ($\vartriangle$) of Fig.
\ref{fig:meanTE-d} represent the transfer entropy from shuffled
data. As expected, the transfer entropy from shuffled data is
smaller than that from the original data, and also the difference
between $T_{I \rightarrow S}$ and $T_{S \rightarrow I}$ is
disappeared below $d \approx 0.02$ and above $d \approx 0.04$ in
both indices. In the range from around 0.02 to around 0.04, number
of states for the indices is 1, while it is still 3 for individual
stocks. Therefore, this difference between them triggers
discrepancy of the transfer entropy between the indices and
stocks.



\begin{figure}[tbph]
\begin{center}
\mbox{
                {\scalebox{0.35}
       {\hspace{0mm}\epsfig{file=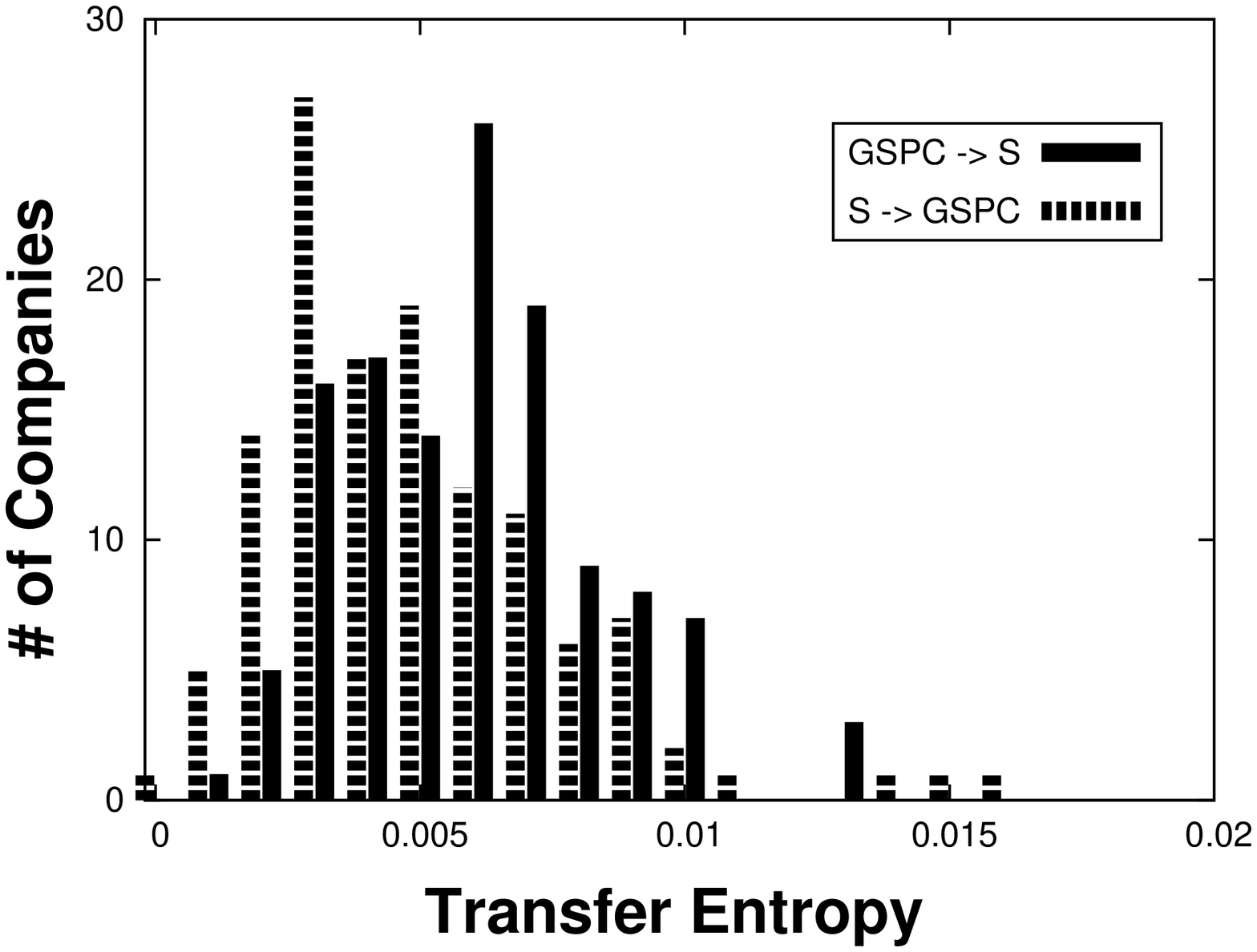, angle=0}} }
         {\scalebox{0.35}
       {\hspace{0mm}\epsfig{file=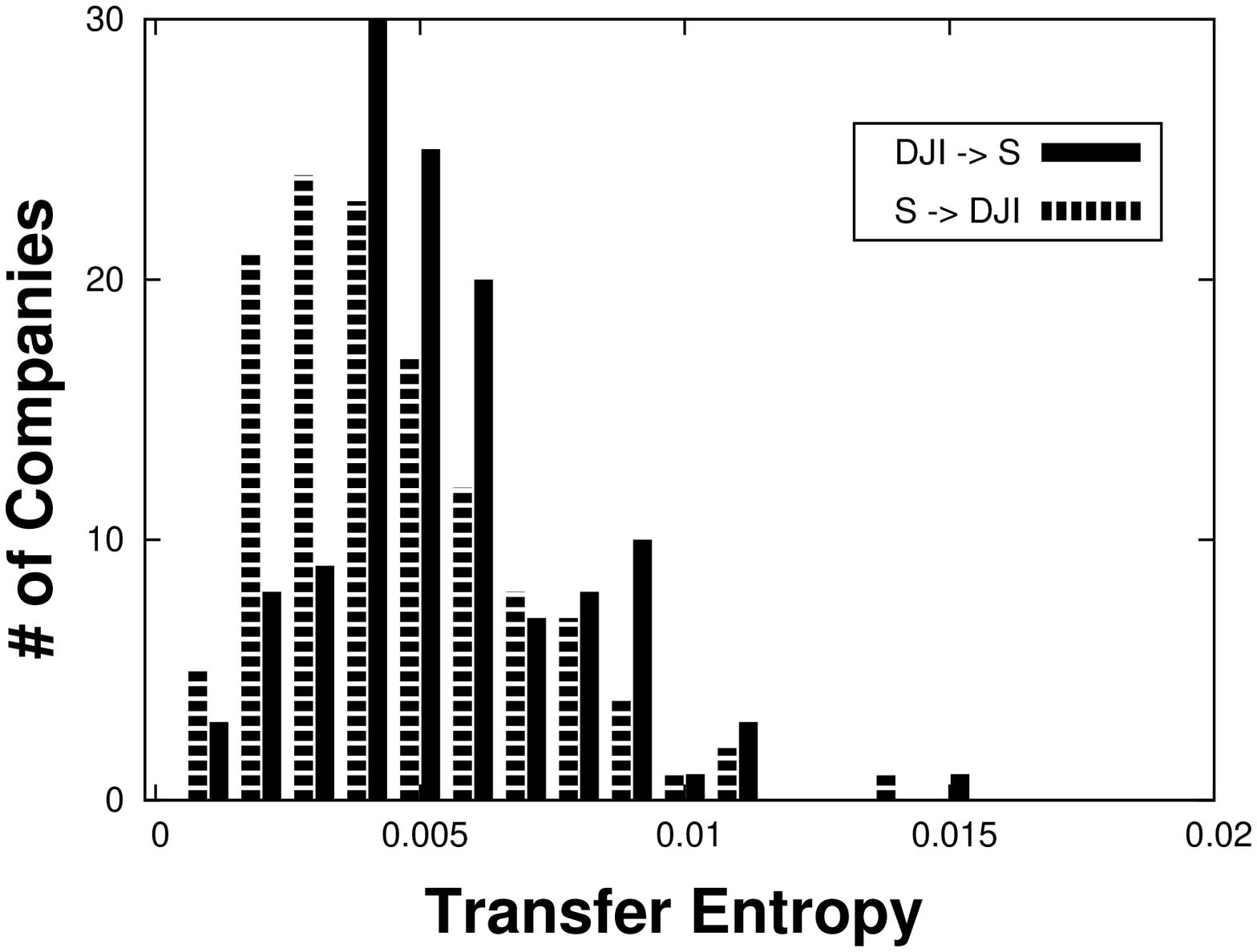, angle=0}} }

} \mbox{
                {\scalebox{0.35}
       {\hspace{0mm}\epsfig{file=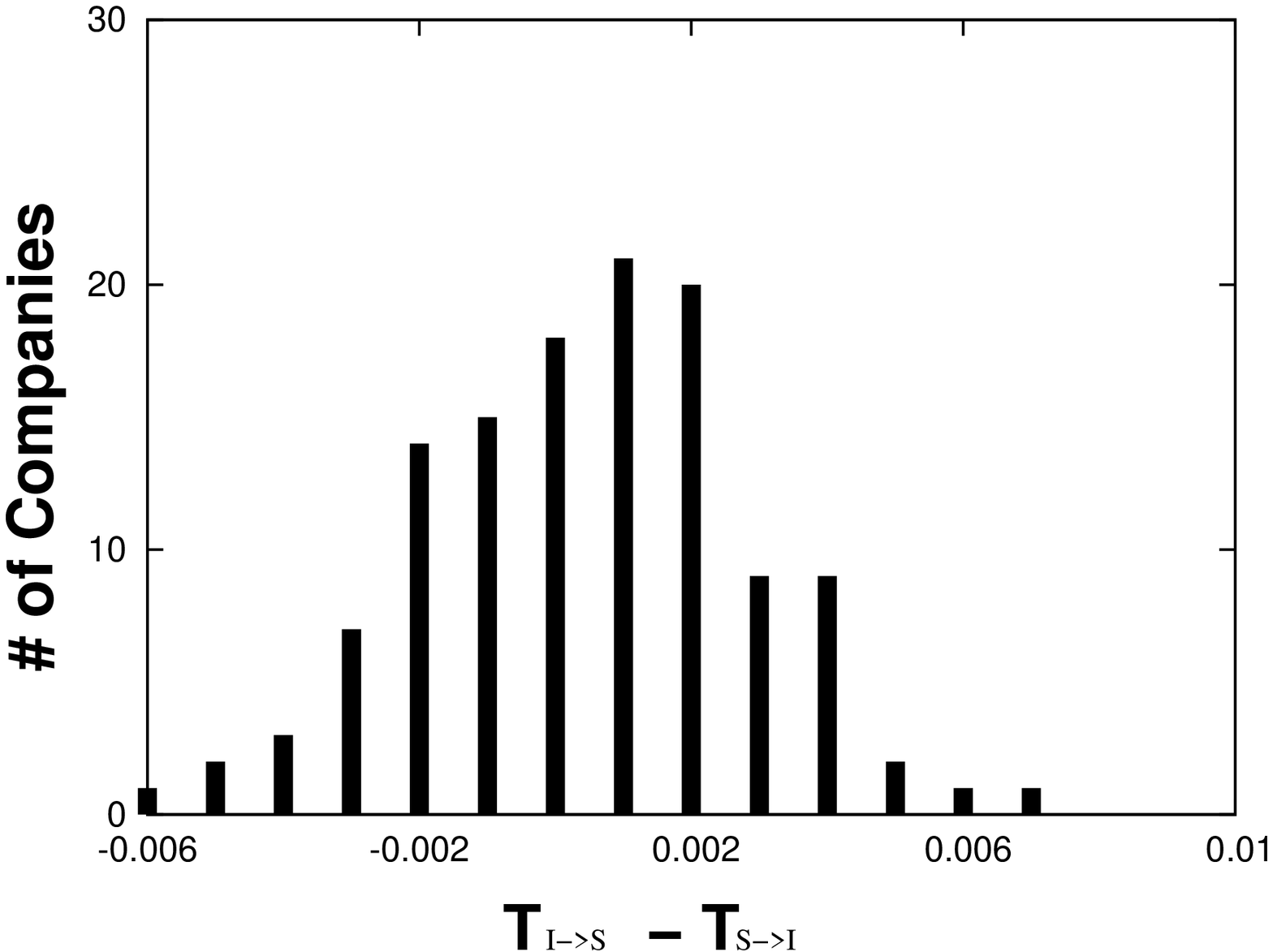, angle=0}} }
         {\scalebox{0.35}
       {\hspace{0mm}\epsfig{file=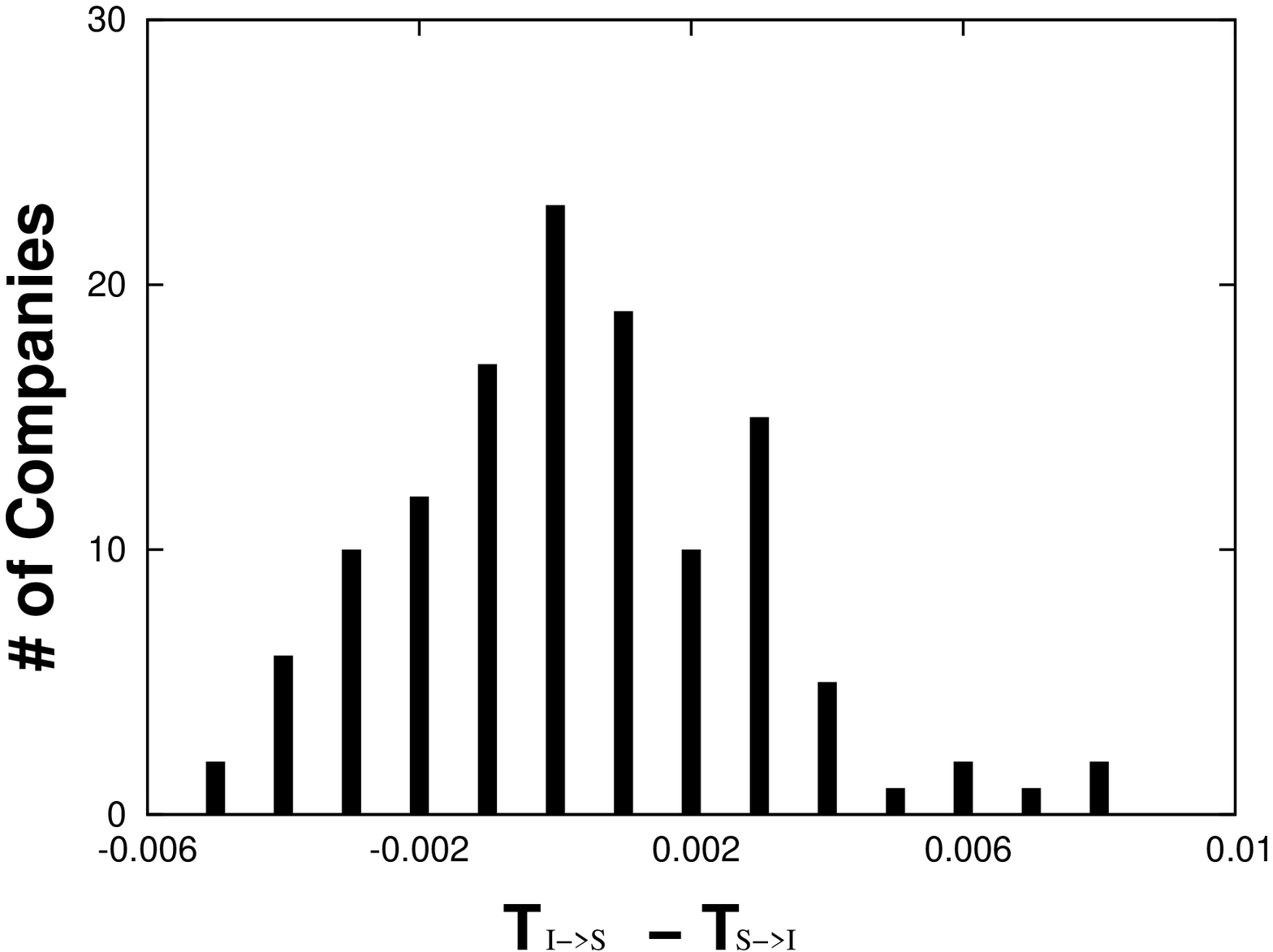, angle=0}} }

}
\end{center}
\vspace{0mm} \caption{At $d=0.015$, the frequency of $T_{I
\rightarrow S}$ and $T_{S \rightarrow I}$ for (a) the GSPC and (b)
the DJI, and the frequency of difference between $T_{I \rightarrow
S}$ and $T_{S \rightarrow I}$ for (c) the GSPC and (d) the DJI.
} \label{fig:dTEstocks}
\end{figure}

Figs. \ref{fig:dTEstocks}(a) and \ref{fig:dTEstocks}(b) show the
frequency of the transfer entropy between composite stock index
and stock prices at $d=0.015$. Frequency distribution of the
transfer entropy from index to stocks is more skewed to right than
that from stocks to index. Figs. \ref{fig:dTEstocks}(c) and
\ref{fig:dTEstocks}(d) show the difference between $T_{I
\rightarrow S}$ and $T_{S \rightarrow I}$. For the majority of
companies, the transfer entropy from index to stocks are larger
than the transfer entropy for the reverse. However, about 35\%
companies gives information to index of the next day.


\begin{figure}[tbph]
\begin{center}
\mbox{
                {\scalebox{0.40}
       {\hspace{0mm}\epsfig{file=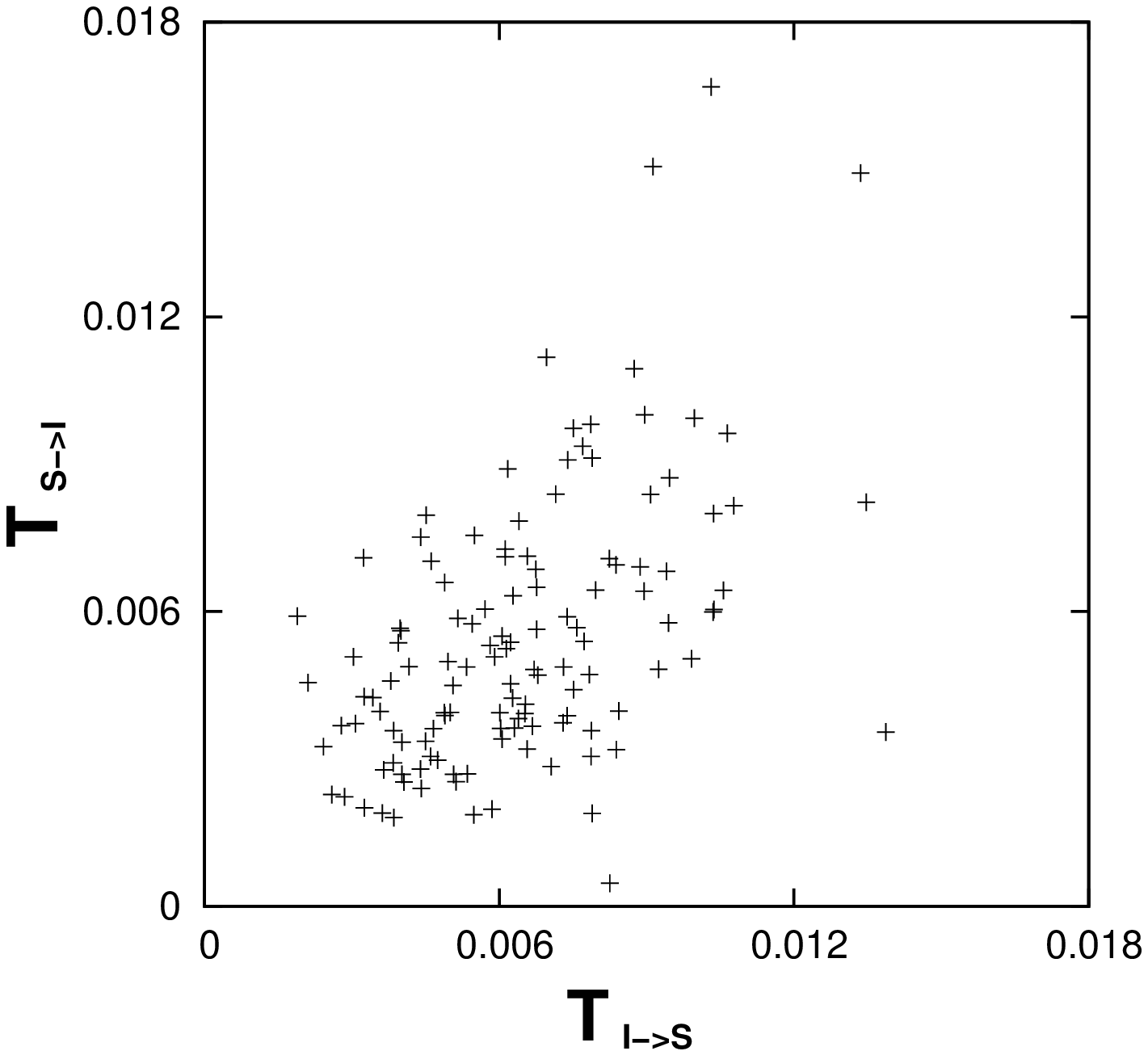, angle=0}} }
         {\scalebox{0.40}
       {\hspace{0mm}\epsfig{file=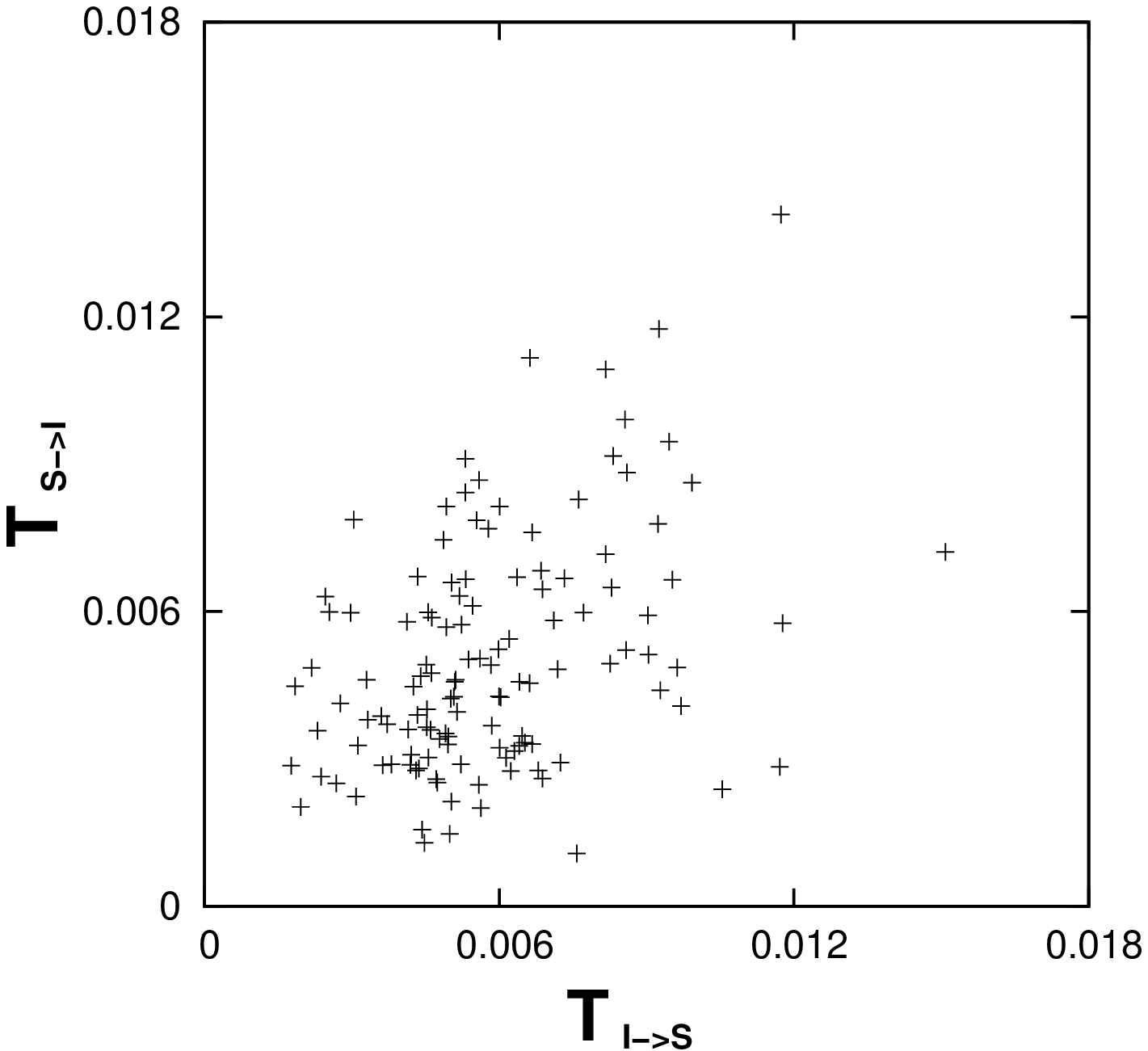, angle=0}} }
}
\end{center}
\vspace{0mm} \caption{The relation between $T_{I \rightarrow S}$
and $T_{S \rightarrow I}$ for (a) the GSPC and (b) the DJI.}
\label{fig:relation}
\end{figure}
\begin{table}[tbp]
\centering%
\begin{tabular}{lllll}
\hline
&GSPC $\rightarrow$ stock & stock $\rightarrow$ GSPC \\
\hline\hline
1&Pepsico Inc.&Centerpoint Energy Inc.\\
2&FPL Group Inc.&Duke Energy Corp.\\
3&Xerox Corp.&Xerox Corp.\\
4&Entergy Corp.&Bristol-Myers Squibb Co.\\
5&Consolidated Edison Inc.&International Business Machines Corp.\\
6&Walt Disney co.&American Electric Power Co. Inc.\\
7&Union Pacific Corp.&PG \& E Corp.\\
8&United Technologies Corp.&TXU Corp.\\
9&Clorox Co.&Wyeth\\
10&Centerpoint Energy Inc.&Consolidated Edison Inc.\\
\hline
&\\
\hline
&DJI $\rightarrow$ stock & stock $\rightarrow$ DJI\\
\hline\hline
1&Walt Disney Co.&Xerox Corp.\\
2&Consolidated Edison Inc.&Centerpoint Energy Inc.\\
3&Xerox Corp.&Willams Companies Inc.\\
4&Whirlpool Corp.&Duke Energy Corp.\\
5&Pepsico Inc.&Southern Co.\\
6&FPL Group Inc.&PG \& E Corp.\\
7&Coca-Cola Co.&American Electric Power Co. Inc.\\
8&United Technologies Corp.&Honeywell International Inc.\\
9&Corning Inc.&Entergy Corp.\\
10&PG \& E Corp.&Bristol-Myers Squibb Co.\\
\hline
\end{tabular}%
\vspace{5mm} \caption{The top 10 companies of the transfer
entropy.} \label{table:company}
\end{table}
Fig. \ref{fig:relation} shows the positive relation between $T_{I
\rightarrow S}$ and $T_{S \rightarrow I}$. The value of
correlation between them is 0.51(9) for the GSPC and 0.40(9) for
the DJI. In Table \ref{table:company}, the top 10 companies of the
transfer entropy is listed. Among the top 10 companies, Xerox
Corp., Entergy Corp., Consolidated Edison Inc., Centerpoint Energy
Inc., and PG \& E belong to the top 10 companies for both $T_{I
\rightarrow S}$ and $T_{S \rightarrow I}$. Both Fig.
\ref{fig:relation} and Table \ref{table:company} show that the
higher $T_{I \rightarrow S}$, the higher $T_{S \rightarrow I}$,
though the average value of $T_{I \rightarrow S}$ is higher than
one of $T_{S \rightarrow I}$. Consequently, individual stocks are
able to be divided into highly connected stocks and lowly
connected stocks to the market.





\section{Conclusion}

The concept of the transfer entropy has been proposed for finding
direction of casuality. Using the measure, we are able to
investigate the information flow between stock index and
individual stocks. Our results indicate that there is a stronger
flow of information from the stock index to the individual stocks
than vice versa, and the transfer entropy for both direction has
positive correlation. Moreover, we expect similar result to the
U.S. market for other stock markets. As a matter of fact, the
result of the information flow in Japan stock market also produces
the same directional casuality although it is not shown in this
paper.

We have desire to find the correlations between the direction of
information flow and company profile. However, we could not find
it yet. The division of individual stocks due to the direction of
casuality between composite stock index and companies may be
useful for the stock investment strategies.

\begin{ack}

This work is supported in part by Creative Research Initiatives of
the Korea Ministry of Science and Technology (O.K.), the Second
Brain Korea 21 project (J.-S.Y.), and Grant No.
R01-2004-000-10148-1 from the Basic Research Program of KOSEF
(J.-S.Y.).

\end{ack}


\end{document}